\documentclass[twocolumn,showpacs,preprintnumbers,amssymb,pra]{revtex4}
\usepackage{pgfplots}
\usepackage{latexsym,epsfig}
\usepackage{graphicx}
\usepackage{dcolumn}
\usepackage{amsthm,amsmath}

\begin{document}
\title{Improved limits on the hadronic and semi-hadronic CP violating parameters and role of a dark force carrier in the electric dipole 
moment of $^{199}$Hg}

\author{B. K. Sahoo \footnote{Email: bijaya@prl.res.in}}
\affiliation{Physical Research Laboratory, Navrangpura, Ahmedabad-380009, India}

\date{Received date; Accepted date}

\begin{abstract}
Combining the recently reported electric dipole moment (EDM) of $^{199}$Hg atom due to breaking of parity and time-reversal symmetries
with the improved relativistic atomic calculations, precise limits on the tensor-pseudotensor (T-PT) electron-nucleus (e-N) coupling 
coefficient and the nuclear Schiff moment (NSM) interactions are determined. Using these limits with the nuclear calculations, we infer 
limits on the EDMs of neutron and proton as $d_n  <  2.2 \times 10^{-26} |e| \rm{cm}$ and $d_p <  2.1 \times 10^{-25} |e| \rm{cm}$, 
respectively, and on the quantum chromodynamics (QCD) parameter and the combined up- and down- quark chromo-EDMs as $|\bar{\theta}| < 
1.1 \times 10^{-10}$ and $|\tilde{d}_u - \tilde{d}_d| < 5.5  \times 10^{-27} |e|\rm{cm}$, respectively. These are the best limits till 
date to probe new sources of CP violation beyond the standard model (SM) from a diamagnetic atom. Role of considering a capable many-body 
method to account the electron correlation effects to all orders for inferring the above limits has been highlighted. From this analysis, 
constraints on the T-PT e-N coupling coefficient with a large range of mass of a possible dark matter carrier $\chi$ between the atomic 
electrons and nucleus are given. 
\end{abstract}

\pacs{21.10.Ky, 31.15.aj, 31.30.Gs, 32.10.Fn}
\maketitle

\section{Introduction}

Observation of permanent electric dipole moment (EDM) in a non-degenerate physical system like an atom is a signature of violations of
parity (P) and time-reversal (T) symmetries \cite{ramsey1,landau,fortson}. As a consequence of CPT theorem \cite{luders}, T violation implies
violation of combined charge conjugation (C) and P symmetries (CP violation). On many counts, investigating CP violation is a subject of 
fundamental importance \cite{khriplovich,roberts}, which motivates to study CP violating interactions extensively on many different systems 
\cite{baker1,hudson,baron,graner,bishof}. It has so far been observed only in the K \cite{christenson} and B mesons \cite{abe,aubert,aaij} at 
the elementary particles level and the results of these experiments are in agreement with the predictions of the Standard Model (SM). 
CP-violation in the SM arises through a complex phase parameter, $\delta$, of the Cabibbo-Kobayashi-Maskawa (CKM) matrix, and due to the 
P,T-odd interactions between the quarks and gluons characterized by the quantum chromodynamics (QCD) parameter $\bar{\theta}$ \cite{ramsey,barr,
pospelov}. But the amount of CP violation described by the SM are not sufficient to explain the observed finiteness of neutrino masses, the 
matter-antimatter asymmetry in the Universe, the existence of dark matter etc. \cite{dine,canetti}. To answer to these profound fundamental 
questions, many extensions of the SM, like the multi-Higgs, supersymmetry, left-right symmetric etc. models have been propounded in the 
literature \cite{barr,pospelov,fukuyama}. Direct probe of new sources of CP violations to plausibly explain many of the above beyond SM 
(BSM) physics can only be carried out at very high energy scale, but it is beyond the reach of the presently operating accelerator 
facilities like large hadron collider (LHC). Therefore, it is imperative to obtain model independent CP violating parameters through indirect 
approaches to validate models supporting the BSM physics \cite{barr,pospelov}. In this context, inferring precise limits on CP violating 
parameters from the electric dipole moments (EDMs) of composite systems such as atoms can be very useful.  

Elementary particles such as electrons and quarks can acquire finite dimensions due to CP violating interactions \cite{barr,fukuyama}. These 
dimensions can be estimated from the knowledge of EDMs of the respective particles. SM  predicts very small EDMs of electrons and quarks 
\cite{pospelov,fukuyama}, but many popular BSM models such as theories like minimal supersymmetric extension of the SM and weak scale 
supersymmetry model (two variants of supersymmetry), multi-Higgs model, two-loop radiative correction to SM etc. \cite{pospelov,barr,barr1,
pospelov1,engel} predict numerous amount of CP violation. Similarly, the coupling coefficients of the CP violating interactions between these 
elementary particles are estimated to be larger in these models than the SM. The effects due to EDMs of the constituent particles and CP violating
interactions among the electrons and quarks are enhanced enormously in the atoms and molecules due to the relativistic and correlation effects 
observed by the bound electrons \cite{ramsey,sandars,ginges}. This makes atoms and molecules are the paramount platforms for investigating CP 
violating effects arising due to BSM. By combining measured EDMs of atoms or molecules with the corresponding calculations it is possible to 
infer information on the EDMs of the electron and quarks and model independent electron-quark (e-q) CP violating coupling coefficients from the 
electron-nucleus (e-N) coupling coefficients deducing them via the electron-nucleon (e-n) couplings \cite{ginges}. Categorically, atoms or 
molecules with the open-shell structure (paramagnetic systems) are sensitive to dominant contributions from the electron EDM ($d_e$) and the 
pseudoscalar-scalar (PS-S) contribution from the CP violating e-N interactions while the closed-shell (diamagnetic) atomic systems are the 
suitable candidates for probing chromo-EDMs of up ($\tilde{d}_u$) and down ($\tilde{d}_d$) quarks, extracting values of the $\bar{\theta}$ 
parameter, the tensor-pseudotensor (T-PT) and scalar-pseudoscalar (S-PS) CP violating coupling e-q coefficients etc. In this work, we 
intend to investigate EDM of the $^{199}$Hg diamagnetic atom to estimate precise limits on different CP violating parameters, which arises 
predominantly due to the nuclear Schiff moment (NSM) \cite{ginges} and the T-PT and S-PS e-N interactions \cite{barr,martensson}. The former 
is associated with the CP violating nucleon-nucleon (n-n) interactions or the EDMs of the proton ($d_p$) and the neutron ($d_n$), which in turn 
originates from the CP violating quark-quark (q-q) interactions as well as EDMs ($d_q$) and chromo-EDMs ($\tilde{d}_q$) of the quarks at the 
elementary particle level \cite{pospelov,engel,dmitriev}, while the later comes from the CP violating e-q interactions.

Though there has not been observation of finite EDM in any system till today, but the on-going atomic and sub-atomic experiments are 
continuously improving limits on the EDMs of neutron \cite{pendelbury}, atoms \cite{graner,bishof} and molecules \cite{hudson,baron}. 
Enhancements of P,T-odd interactions in atoms and molecules are larger than neutron with respect to the fundamental particles. Molecules
are the better candidates for finding out more precise limit on $d_e$ owing to strong polarization effects exhibited within the 
paramagnetic polar molecules that produce large internal electric fields and causes large energy shifts in the molecular energy levels
than the atomic energy levels when $d_e$ of the valence electron interacts with an external electric field. However, consideration of diamagnetic atoms than diamagnetic molecules are 
advantageous for studying the NSM and T-PT e-N interactions as atomic systems involve comparatively simpler nuclear interactions than molecules.
It to be noted that for deducing e-q interaction coefficients from the e-N interactions, accurate knowledge of nuclear interactions are 
also essential \cite{ginges,engel,dmitriev}. So far, measurements only in the $^{129}$Xe \cite{rosenberry}, $^{199}$Hg \cite{graner,griffith} and 
$^{225}$Ra \cite{bishof,parker} diamagnetic atoms have been performed among which the EDM of the $^{199}$Hg atom is reported more 
precisely from a recent measurement as $d( ^{199}\rm{Hg})=(-2.20 \pm 2.75(\rm{stat}) \pm 1.48 (\rm{sys})) \times 10^{-30}$ $e$cm 
corresponding to an upper limit  $|d( ^{199}\rm{Hg})| < 7.4 \times 10^{-30}$ $e$cm with 95\% confidence level \cite{graner}. In our 
previous work \cite{yashpal}, we had obtained limits on $\bar{\theta}$ and $|\tilde{d}_u- \tilde{d}_d|$ by combining atomic calculations, 
carried out by employing a relativistic coupled-cluster (RCC) method, with the measurement on EDM of $^{199}$Hg reported earlier in Ref. 
\cite{griffith}. We had also demonstrated in that work that other relativistic methods such as random phase approximation (RPA) estimate 
very large magnitudes of atomic results for neglecting some of the important electron correlation effects like pair-correlation effects. 
Hence, use of those calculations give rise much lower limits on the above quantities. We attempt to review these limits here by 
considering the latest experimental data from Ref. \cite{graner} and embodying more contributions in the atomic calculations from higher 
order electron correlations, especially from the pair-correlation effects, through the RCC method. Moreover, the required  nuclear 
calculations have also been revised extensively all along \cite{dekens,vries}.

In fact, observation of dark matter in the universe has unearthed another Pandora's box to study interactions among the particles in their 
presence. It is believed that matters could interact with each other in the presence of dark forces by exchanging gauge boson types of 
particles, known as dark force carriers, possessing very weak couplings. It is again strenuous to observe these effects through the 
accelerator based methods, even though their masses can be of reasonable size, because of their exceptionally smaller cross-sections. In a 
recent work \cite{derevianko}, the P,T-odd interaction between the electrons and nucleons in $^{199}$Hg due to the exchange of a massive 
light gauge boson $\chi$ was investigated and constraint on the T-PT e-N coupling constant with a range of mass of $\chi$ ($m_{\chi}$) 
was given. In that study, the previously measured experimental data on $^{199}$Hg EDM from Ref. \cite{griffith} was used, which has now 
been improved almost by one order very recently \cite{graner}, and a relativistic RPA method, accounting only the core-polarization 
effects of electron correlations, was employed to estimate the T-PT interaction enhancement due to the exchange of $\chi$ between the 
electrons and nucleus in this atom. We improve accuracies in the atomic calculations further using the RCC method and use the latest 
experimental result to put more stringent limit on the T-PT e-N coupling constant. 

In order to ascertain validity of the employed methods in the present study, we also perform calculation of the dipole polarizability 
($\alpha_d$) of Hg atom, which has similar expression for evaluating the P,T-odd interactions, using the considered many-body methods 
and compare the results with the available experimental value.

\begin{table}[t]
\caption{\label{tab1} 
Calculated $\alpha_d$ value in $|e|a_0^2$ and $d_a$ values with the T-PT e-N interaction (given in $10^{-20}C_T \langle \sigma_N\rangle$
$|e|$cm) with $m_{\chi} = \infty$ and nuclear Fermi charge distribution and the NSM interaction (given in ($10^{-17}[S/|e|fm^3]$ $|e|$cm)
in the $^{199}$Hg atom from the DHF, MBPT(2), MBPT(3), RPA and CCSD methods. Good agreement between the experimental and CCSD values of 
$\alpha_d$ suggests, CCSD method is capable of producing more accurate results.} 
\begin{ruledtabular}
\begin{tabular}{lccc}
  &     &   \\
Method & $\alpha_d$ & T-PT  & NSM  \\
\hline 
   &  &  & \\
DF & 40.95 & $-$2.39 & $-$1.20  \\ 
MBPT(2)& 34.18 & $-$4.48  &$-$2.30 \\
MBPT(3)& 22.98 & $-$3.33  &$-$1.72 \\
RPA    & 44.98 & $-$5.89  &$-$2.94 \\
CCSD & 34.51 & $-3.17$ &  $-1.76$ \\
\hline
Experiment \cite{goebel} & 33.91(34) &  & \\
\end{tabular}
\end{ruledtabular}   
\end{table}

\section{Theoretical formalism}

The P,T-odd Lagrangian for the e-n interaction is given by \cite{pospelov}
\begin{eqnarray}
\mathcal{L}_{e-n}^{PT} &=& C_T^{e-n} \varepsilon_{\mu \nu \alpha \beta} \bar{\psi}_e \sigma^{\mu \nu} \psi_e  
\bar{\psi}_n \sigma^{\alpha \beta} \psi_n \nonumber \\ && + C_P^{e-n}  \bar{\psi}_e  \psi_e \ \bar{\psi}_n i \gamma_5 \psi_n,
\end{eqnarray}
where $\varepsilon_{\mu \nu \alpha \beta}$ is the Levi-Civita symbol and $\sigma_{\mu \nu} = \frac{i}{2}[\gamma_\mu, \gamma_\nu]$ with
the $\gamma$'s being the Dirac matrices. $C_T^{e-n}$ and $C_P^{e-n}$ are the T-PT and S-PS e-n interaction coupling constants respectively. 
Here $\psi_n$ and $\psi_e$ represent the Dirac wave functions of the nucleon and electron respectively. Assuming that the T-PT and S-PS e-N 
interactions act independently, we can consider them individually in the atomic calculations for which the respective e-N interaction 
Hamiltonians are given as \cite{ginges,martensson,dzuba}
\begin{eqnarray}
 && H_{e-N}^{TPT} = i \sqrt{2} G_F C_T \sum_e \mbox{\boldmath $\sigma_N \cdot \gamma$} \rho_N(r) \label{htpt}\\
\text{and} && \nonumber \\
&& H_{e-N}^{SPS} = - \frac{G_F}{\sqrt{2}m_n c} C_P \sum_e \gamma_0 \mbox{\boldmath $\sigma_N \cdot \nabla$} \rho_N(r), 
\end{eqnarray}
where $G_F$ is the Fermi constant, $C_T$ and $C_P$ are the T-PT and S-PS e-N coupling constants, {\boldmath$\sigma_N$}$=\langle 
\sigma_N \rangle {\bf I}/I$ is the Pauli spinor of the nucleus with spin $I$, $\rho_N(r)$ is the nuclear density, $m_n$ is the 
nucleon mass and $c$ is the speed of light. $C_P$ can be estimated with reasonable accuracy from the knowledge of $C_T$ using an 
empirical relation \cite{ginges,dzuba}
\begin{eqnarray}
C_P \approx 3.8 \times 10^3 \times \frac{A^{1/3}}{Z} C_T
\label{eqcp}
\end{eqnarray}
with the atomic number $Z$ and mass $A$. We are only interested in determining the $C_T$ value here so that $C_P$ can be estimated 
using the above relation.

The Lagrangian for the P,T-odd pion-nucleon-nucleon ($\pi$-n-n) interaction that also contribute predominantly to the EDMs of the 
diamagnetic atoms is given by \cite{pospelov}
\begin{eqnarray}
\mathcal{L}^{\pi n n}_{e-n} &=& \bar{g}_0 \bar{\psi}_n \tau^i \psi_n \pi^i + \bar{g}_1 
\bar{\psi}_n \psi_n \pi^0 \nonumber \\ && + \bar{g}_2 \big ( \bar{\psi}_n \tau^i \psi_n \pi^i - 
3 \bar{\psi}_n \tau^3 \psi_n \pi^0 \big )
\end{eqnarray}
where the couplings $\bar{g}_i$ with the superscript $i=$ 1, 2, 3 represent the isospin components. The corresponding e-N interaction 
Hamiltonian is given by \cite{ginges,dzuba}
 \begin{eqnarray}
  H_{e-N}^{NSM}= \frac{3{\bf S.r}}{B_4} \rho_N(r),
 \end{eqnarray}
where ${\bf S}=S \frac{{\bf I}}{I}$ is the NSM and $B_4=\int_0^{\infty} dr r^4 \rho_N(r)$. The magnitude of the NSM $S$
can be expressed as \cite{ginges,engel}
\begin{eqnarray}
S = g_{\pi n n} \times (a_0 \bar{g}_{\pi n n}^{(0)} + a_1 \bar{g}_{\pi n n}^{(1)} + a_2 \bar{g}_{\pi n n}^{(2)}),
\end{eqnarray}
where $g_{\pi nn} \simeq 13.5$ is the CP-even $\pi$-n-n coupling constant, $a_i$s are the polarizations of the nuclear charge 
distribution that can be computed to reasonably accuracy using the Skyrme effective interactions in the Hartree-Fock-Bogoliubov 
mean-field method \cite{engel} and $\bar{g}_{\pi n n}^{(i)}$s with $i=$ 1, 2, 3 representing the isospin components of the CP-odd 
$\pi$-n-n coupling constants. These couplings are related to $\tilde{d}_u$ and $\tilde{d}_d$ as $\bar{g}_{\pi n n}^{(1)} \approx 2
\times  10^{-12} \times (\tilde{d}_u - \tilde{d}_d)$ \cite{pospelov,pospelov1,vries} and $\bar{g}_{\pi n n}^{(0)}/
\bar{g}_{\pi n n}^{(1)} \approx 0.2 \times (\tilde{d}_u + \tilde{d}_d)/(\tilde{d}_u -\tilde{d}_d)$ \cite{pospelov,dekens}, where chromo-EDMs
of quarks are scaled with $\times 10^{-26} |e| \rm{cm}$. It is 
also related to $\bar{\theta}$ parameter by $|\bar{g}_{\pi n n}^{(1)}|=0.018(7) \bar{\theta}$ \cite{dekens}. From the nuclear 
calculations, one can obtain $S \simeq (1.9d_n+0.2d_p)$ fm$^2$ \cite{dmitriev}.

Again allowing the exchange of dark matter gauge boson $\chi$ between the electrons and nucleus in the considered atoms, the T-PT
e-N interaction Hamiltonian can be expressed by \cite{derevianko}
\begin{eqnarray}
 H_{e-N}^{TPT} = i \sqrt{2} G_F C_T^{\chi} \sum_e \mbox{\boldmath $\sigma_N \cdot \gamma$} \rho_N^{\chi}(r), 
\end{eqnarray}
where $C_T^{\chi}$ and $\rho_N^{\chi}(r)$ are the corresponding modified T-PT coupling constant and nuclear density, respectively,
with the exchange of dark force carrier $\chi$. It to be noted that the above expression is the most general one in describing the contact 
interaction due to finiteness of the exchange boson $m_{\chi}$ and Eq. (\ref{htpt}) can be treated as the special case valid in the 
limit $m_{\chi} \rightarrow \infty$. Therefore, it yields $C_T^{\chi}\xrightarrow{m_{\chi}\rightarrow \infty} C_T$ and $\rho_N^{\chi}
\xrightarrow{m_{\chi}\rightarrow \infty} \rho_N$. Owing to this fact, it is needed to define a contact potential of Yukawa-type 
between the electrons and nucleons in an atom due to exchange of $m_{\chi}$ such as \cite{derevianko}
\begin{eqnarray}
 V_{\chi}(r,r')= \frac{e^{-m_{\chi}c|{\bf r}-{\bf r}'|}}{4 \pi |{\bf r}-{\bf r}'|} ,
\end{eqnarray}
where $r$ and $r'$ represent coordinates of the electrons and the nucleons respectively. It can be shown that in the large mass limit, 
this Yukawa-type potential behaves like $\delta^3(r-r')/(m_{\chi}c)^2$. Considering this potential, $\rho_N^{\chi}(r)$ is 
given by \cite{derevianko}
\begin{eqnarray}
 \rho_N^{\chi}(r)= 4 \pi (m_{\chi}c)^2 \int dr' V_{\chi}(r,r') \rho_N(r') .
\end{eqnarray}
For convenient determination of $V_{\chi}(r,r')$, uniform charge density of the nucleus is taken into account in which we define
\begin{eqnarray}
 \rho_N(r) = \rho_0 \Theta(1-(r/R)),
\end{eqnarray}
with $\rho_0=\frac{3 Z}{4 \pi R^3}$ for the radius of the sphere $R$ within which the nucleons are uniformly distributed and 
$\Theta$ is the Heaviside step function. Using this approximation, it yields \cite{derevianko} 
\small{
\begin{flushleft}
\begin{eqnarray}
\rho_N^{\chi}(r)  = \rho_0 \frac{\lambda_{\chi}}{r} 
\left\{\begin{array}{rl}
 \frac{r} {\lambda_{\chi}} - e^{-\frac{R}{\lambda_{\chi}}} \left ( 1+\frac{R}{\lambda_{\chi}} \right ) \sinh \left ( \frac{r}{\lambda_{\chi}} \right ),  & \mbox{$r \leq R$}\\
   e^{-\frac{r}{\lambda_{\chi}}} \left (\frac{R}{\lambda_{\chi}} \cosh \left ( \frac{R}{\lambda_{\chi}} \right ) - \sinh \left ( \frac{R}{\lambda_{\chi}} \right ) \right ),  & \mbox{$r>R$,}
\end{array}\right.  \nonumber \\ 
\label{eq12}
\end{eqnarray}
\end{flushleft}}
where $\lambda_{\chi}=\hbar/(m_{\chi}c)$. As can be seen this density functional goes as $\rho_N^{\chi}(r) \propto 
\frac{e^{-r/\lambda_{\chi}}}{r}$. In order to demonstrate the validity of considering nuclear uniform charge distribution in our 
EDM calculation, we also compare the values with the result obtained by considering the nuclear Fermi charge density distribution 
for the limit $m_{\chi} \rightarrow \infty$ as
\begin{equation}
\rho_N(r)=\frac{\rho_{0}}{1+e^{(r-b)/a}},
\end{equation}
where $\rho_0$ is the corresponding normalization constant, $b$ is the half-charge radius and $a= 2.3/4(ln3)$ is related to the 
skin thickness. 

Denoting $d_a^{\chi}$ and $d_a^c$ are the EDM of the atom due to exchange of finite mass of $m_{\chi}$ (due to contact interaction) and 
with infinite mass of $m_{\chi}$ respectively, we can express a ratio as
\begin{eqnarray}
 {\cal R}^{\chi} = \frac{d_a^{\chi}/C_T^{\chi} }{d_a^c/C_T} ,
\end{eqnarray}
so that it follows
\begin{eqnarray}
 d_a^{\chi} \equiv C_T^{\chi} \left ( \frac{d_a^c}{C_T} \right ) {\cal R}^{\chi} \le \text{Measured EDM of the atom}.
\end{eqnarray}
Thus, using the experimental EDM value of $^{199}$Hg we can express
\begin{eqnarray}
 |C_T^{\chi}| \le |d( ^{199}{\rm{Hg}})| \left | \frac{C_T}{d_a^c} \frac{1}{{\cal R}^{\chi}} \right | .
 \label{eqctx}
\end{eqnarray}
By calculating $\frac{C_T}{d_a^c}$ for infinite value of $m_{\chi}$ and ${\cal R}^{\chi}$ for a wide range of $m_{\chi}$, we shall 
be able to constraint on $C_T^{\chi}$ using the experimental value of $d( ^{199}{\rm{Hg}})$. For this purpose, it is necessary to evaluate 
$\frac{C_T}{d_a^c}$ and ${\cal R}^{\chi}$ reliably by employing potential relativistic atomic many-body theories such as RPA and RCC methods 
that are described briefly in the following section.

\begin{figure}[t]
\includegraphics[width=9.0cm,height=6.0cm]{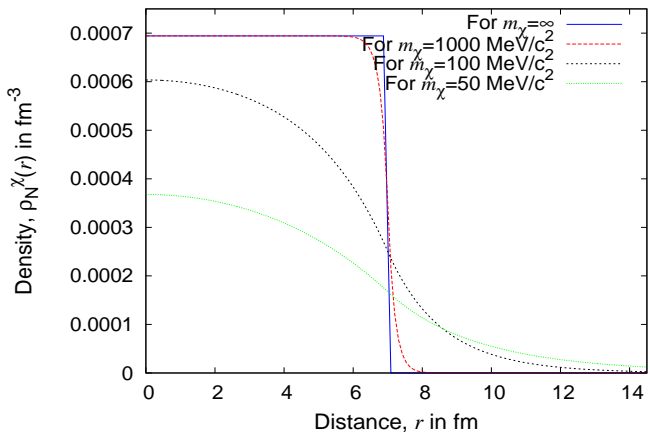}
\caption{(Color online) Behavior of $\rho_N^{\chi}$ values with the distance from the origin of the nucleus of $^{199}$Hg in the 
uniform charge distribution for different values of $m_{\chi}$ (in MeV/$c^2$. This is perfectly in agreement with the findings
of Ref. \cite{derevianko}.}
 \label{fig1}
 \end{figure}

\section{Method of calculations}

To pursue calculations of atomic wave functions, we start with the mean-field wave function ($|\Phi_0 \rangle$) of the 
ground state of $^{199}$Hg using the Dirac-Hartree-Fock (DHF) method in the Dirac-Coulomb (DC) atomic Hamiltonian. With the reference 
state $|\Phi_0 \rangle$, the ground state ($|\Psi_0 \rangle$) is determined by appending the electron correlations in the RPA and 
RCC methods as described in our previous work \cite{yashpal}. These methods are briefly described below. 

Using the RCC method ansatz, we can express \cite{bartlett} 
\begin{eqnarray}
 |\Psi_0^{(0)} \rangle = e^{T^{(0)}} |\Phi_0 \rangle,
\end{eqnarray}
where $T^{(0)}$ is the even parity RCC operators that takes care of the neglected correlation effects of the DHF method among the electrons 
to all orders by exciting them to the higher states with respect to $|\Phi_0 \rangle$. In the presence of a P,T-odd interaction, this wave 
function is modified to 
\begin{eqnarray}
 |\Psi_0 \rangle = e^T |\Phi_0 \rangle=e^{T^{(0)}+\lambda T^{(1)}},
\end{eqnarray}
where $T=T^{(0)}+\lambda T^{(1)}$ accounts both the electron correlation effects due to the electromagnetic interactions by $T^{(0)}$ and 
electromagnetic interactions along with the P,T-odd weak interaction by $T^{(1)}$ to all orders. Here $\lambda$ can be 
interpreted as the strength of the coupling coefficient of the considered P,T-odd interaction. Note that $|\Psi_0\rangle$ becomes a mixed 
parity state in contrast to the even parity wave function $|\Psi_0^{(0)}\rangle$ owing to addition of the odd-parity P,T-interaction 
Hamiltonians. Since these weak interactions are much smaller than the electromagnetic interactions in the atoms, we can consider the 
electron correlation effects to all orders and the P,T-odd interaction to the first order by approximating as
\begin{eqnarray}
 |\Psi_0 \rangle \simeq |\Psi_0^{(0)} \rangle + \lambda |\Psi_0^{(1)} \rangle,
\end{eqnarray}
with the corresponding weak coupling coefficient $\lambda$ which is either $C_T^{\chi}$ or $S$ for the T-PT and NSM interactions, 
respectively. This yields
\begin{eqnarray}
 |\Psi_0^{(1)} \rangle = e^{T^{(0)}} T^{(1)} |\Phi_0 \rangle.
\end{eqnarray}
It implies $|\Psi_0^{(1)} \rangle$ is an odd parity wave function due to $T^{(1)}$. We solve first the amplitudes of the $T^{(0)}$ 
operators following which we obtain amplitudes of the $T^{(1)}$ operators. For the calculations of both $|\Psi_0^{(0)} \rangle$ and 
$|\Psi_0^{(1)} \rangle$, we allow only all possible singles and doubles excitations by defining $T^{(0)}=T_1^{(0)}+T_2^{(0)}$ and 
$T^{(1)}=T_1^{(1)}+T_2^{(1)}$ in the RCC method (CCSD method). 

 \begin{table}[t]
\caption{\label{tab2} 
Variation in the $d_a^{\chi}$ values (in $10^{-18}C_T^{\chi} \langle \sigma_N\rangle$ $|e|$cm) of $^{199}$Hg with $m_{\chi}$ (in MeV/$c^2$) 
from the DHF, RPA and CCSD methods with uniform nuclear charge distribution.} 
\begin{ruledtabular}
\begin{tabular}{lccc}
$m_{\chi}$ & DHF & RPA  & CCSD  \\
\hline 
  &   & \\
$\infty$ & $-240.33$ & $-$591.36 & $-$318.46  \\ 
4882.81  & $-238.85$  & $-$587.71  & $-316.49$ \\
3906.25  & $-239.06$  & $-588.22$  & $-316.77$ \\
3000.00 & $-239.15$ &  $-588.46$ & $-316.90$ \\
2441.41  & $-239.15$ & $-588.46$  & $-316.90$   \\
1953.13  & $-239.11$ &  $-588.35$ &  $-316.84$ \\
1500.00  & $-239.03$ &  $-588.17$  &  $-316.74$ \\
976.56   & $-238.88$ & $-587.80$  & $-316.54$  \\
488.28   & $-238.38$ & $-586.57$ & $-315.88$ \\
195.31  & $-235.40$ & $-579.18$  & $-311.92$ \\
97.65  & $-227.09$ & $-558.80$ & $-300.89$ \\
70.0 & $-219.43$  & $-539.97$ & $-290.72$ \\
39.06  & $-198.95$ & $-489.63$ & $-263.52$ \\
19.53  & $-165.31$ & $-406.98$ & $-218.86$ \\
7.81 & $-115.75$ & $-285.37$ & $-153.01$ \\
3.0 & $-67.80$ & $-168.13$ & $-89.38$ \\
1.5625  & $-40.17$ & $-101.01$ & $-52.93$ \\
1.2 &  $-30.70$ & $-78.09$ & $-40.54$ \\
1.0  & $-24.87$ & $-64.02$ & $-32.95$ \\
0.9  &  $-21.79$ & $-56.59$ & $-28.96$ \\
0.8  & $-18.62$ & $-48.92$ & $-24.86$ \\
0.7  & $-15.37$ & $-41.07$ & $-20.67$ \\
0.6 & $-12.09$ & $-33.10$ & $-16.46$ \\
0.5 & $-8.85$ & $-25.18$ & $-12.31$ \\
0.4  & $-5.80$ & $-17.56$ & $-8.38$ \\
0.3125 & $-3.45$ & $-11.45$ & $-5.30$ \\
0.25 & $-2.07$ & $-7.63$ & $-3.44$ \\
0.2 & $-1.22$ & $-5.02$ & $-2.22$ \\
0.15 & $-0.62$ & $-2.91$ & $-1.26$ \\
0.1  & $-0.27$ & $-1.34$ & $-0.59$ \\
0.0625 & $-0.13$ & $-0.56$ & $-0.26$ \\
0.0125 & $-0.01$ & $-0.03$ & $-0.02$ \\
0.0025 & $-6.0\times 10^{-5}$ & $-1.2\times 10^{-3}$ & $-6.0\times 10^{-4}$\\
0.00125  & $1.1\times 10^{-5}$ & $-2.4 \times 10^{-4}$ & $-1.3\times 10^{-4}$ \\
0.0008 & $8.0\times 10^{-6}$ & $-8.0\times 10^{-5}$ &  $-5.4\times 10^{-5}$\\
\end{tabular} 
\end{ruledtabular}   
\end{table}

As described in Ref. \cite{yashpal}, these wave functions in RPA are given by
\begin{eqnarray}
 |\Psi_0^{(0)} \rangle \approx |\Phi_0 \rangle \ \ \ \ \text{and} \ \ \ \
 |\Psi_1^{(0)} \rangle \approx \Omega_I^{(\infty,1)} |\Phi_0 \rangle ,
\end{eqnarray}
where $\Omega_I^{(\infty,1)}$ is the RPA excitation operator that accounts only the core-polarization electron correlation effects to all orders 
and the P,T-odd interaction to one order through the single excitations. As were highlighted in our previous works on similar studies 
\cite{yashpal,bijaya,yashpal1}, the electron pair-correlation effects contribute significantly to these quantities and cancel out with the 
core-polarization effects substantially giving rise to smaller values to the final results. Nonetheless, both these core-polarization and pair-correlation 
effects are being incorporated in the equally footing to all orders and are solved self-consistently using the coupled equations in the RCC 
method \cite{yashpal,bijaya,yashpal1}.

\begin{figure}[t]
\includegraphics[width=9.0cm,height=6.5cm]{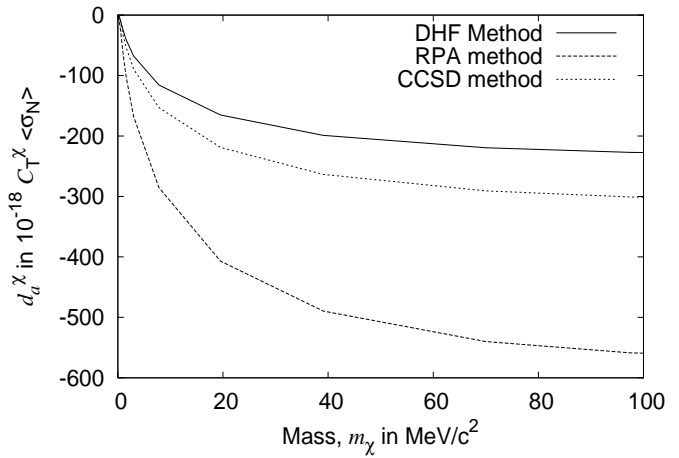}
\caption{(Color online) Demonstration of trends in the $d_a^{\chi} $ values (in $10^{-18}C_T^{\chi} \langle \sigma_N\rangle$ $|e|$cm)
with $m_{\chi}$ (in MeV/$c^2$) from the DHF, RPA and CCSD methods. This clearly shows RPA values are far away from the DHF results, while 
the CCSD results are close to the DHF values. The differences between the RPA and CCSD results are the clear indication of strong 
contributions by the electron pair-correlation effects.}
 \label{fig2}
 \end{figure}

After obtaining the $|\Psi_0^{(0)} \rangle $ and $|\Psi_0^{(1)} \rangle$ wave functions in the above formalisms, we evaluate
\begin{eqnarray}  
\frac{d_a}{\lambda} \equiv \mathcal{Y} &=& 2 \langle \Psi_0^{(0)} | D |\Psi_0^{(1)} \rangle,              
\label{eq5}
\end{eqnarray}
for the atomic EDM $d_a$ that corresponds to $\mathcal{Y}=d_a^{\chi}/{C_T^{\chi}}$ and $\mathcal{Y}=d_a/S$ for the T-PT and NSM 
interactions, respectively. This in the RPA is given by 
\begin{eqnarray}
 \mathcal{Y} &=& 2 \langle \Phi_0 | D \Omega_I^{(\infty,1)} |\Phi_0 \rangle.
\end{eqnarray}
In the RCC framework, we can obtain \cite{yashpal}
\begin{eqnarray}
 \mathcal{Y} &=& 2 \langle \Phi_0 | [ e^{{T^{(0)}}^{\dagger}} D e^{T^{(0)}} T^{(1)} ]_{conn} |\Phi_0 \rangle,
\end{eqnarray}
where the subscript $conn$ implies all the connected terms are evaluated by adopting the Wick's generalized theorem \cite{bartlett}. As can 
be seen, the above RCC expression has a non-truncative series $e^{{T^{(0)}}^{\dagger}} D e^{T^{(0)}}$. By replacing the P,T-odd 
interaction Hamiltonian by a dipole operator $D$ and considering $\lambda=1$ in the evaluation of $|\Psi_0^{(1)} \rangle$, it will
give $\mathcal{Y}$ as $\alpha_d$ in the above expressions. As per the generalized Wick's
theorem, we break this expression into effective one-body and two-body terms for their evaluations. In our previous work \cite{yashpal}, 
we had demonstrated an approach using which the non-truncative effective one-body terms were calculated self-consistently but contributions 
from the effective two-body terms were estimated approximately. In this work, we have included more possible effective two-body terms in a 
self-consistent procedure using the intermediate diagrammatic techniques in the similar manner as discussed in Ref. \cite{bartlett} for 
accurate evaluation of $\mathcal{Y}$. These corrections are part of the pair-correlation effects and need to be determined rigorously for 
attaining more reliable results. Contributions from these diagrams, especially that arise after contracting with the $T_2^{(1)}$ operators, 
are coming out to be whopping amount and cancel out further with the core-polarization contributions. As a matter of fact, we find the 
differences between the RPA and CCSD results are quite significant. We also present results considering only one order Coulomb interaction 
and two orders of Coulomb interactions along with the P,T-odd interaction in the atomic wave function in a perturbative approach of the 
second order many-body perturbation theory (MBPT(2) method) and the third order many-body perturbation theory (MBPT(3) method) respectively. 
In our EDM studies on the $^{223}$Rn \cite{bijaya} and $^{225}$Ra \cite{yashpal1} atoms we had demonstrated that the lowest order 
core-polarization effects enter through the MBPT(2) method while the lowest order pair-correlation effects contribute at the MBPT(3) method.
Thus, it is possible to understand about the importance of the pair-correlation contributions from the differences between the results from 
the MBPT(2) and MBPT(3) methods.

\section{Results and Discussion}

In Table \ref{tab1}, we present the $d_a$ values by calculating $\mathcal{Y}$ for the corresponding T-PT e-N and NSM interactions in the 
$^{199}$Hg atom using relativistic many body methods at different levels of the approximations and considering nuclear Fermi charge 
distribution. The DHF results are obtained using the wave functions from the DHF method. We had already discussed results reported by others 
employing various many-body methods in our previous work \cite{yashpal}. Here we present the improved CCSD results compared to the values 
given in Ref. \cite{yashpal} by taking into account the additional effects mentioned in the previous section. In fact, we had already given 
results from the DHF, MBPT(2), MBPT(3) and RPA methods in Ref. \cite{yashpal} as listed in the above table. However, we present them again 
here to highlight the reasons for which differences between the RPA and CCSD results occur. This could justify the intention for calculating the 
$C_T^{\chi}$ coupling coefficients due to the exchange of dark matter candidate $\chi$ using the CCSD method, which were determined earlier by 
the RPA method \cite{derevianko}. It is worth mentioning here that our DHF and RPA results given in the above table agree well with another
calculation reported in Ref. \cite{dzuba}. Again, $\alpha_d$ values obtained from different methods are given and compared with the 
experimental result in the same table. As can be seen, the CCSD value 34.51 $|e|a_0^2$ is in very good agreement with the experimental value 
33.91(34) $|e|a_0^2$ \cite{goebel}, while RPA estimates its value far away from the experimental result. The reason for which the CCSD values are coming out to be off from the RPA values can be understood 
from the differences in the results of the  MBPT(2) and MBPT(3) methods. From the computational point of view the DHF and MBPT(2) methods take 
only less than an hour to produce the results while it takes few hours to perform calculations using the MBPT(3) and RPA methods to obtain the 
above quantities. However, the bottle-neck of carrying out CCSD calculations is that it takes days to complete a set of results. Thus, the 
improvements in the results by the CCSD method are done with the cost of huge computational resources.

\begin{figure}[t]
\includegraphics[width=9.0cm,height=6.5cm]{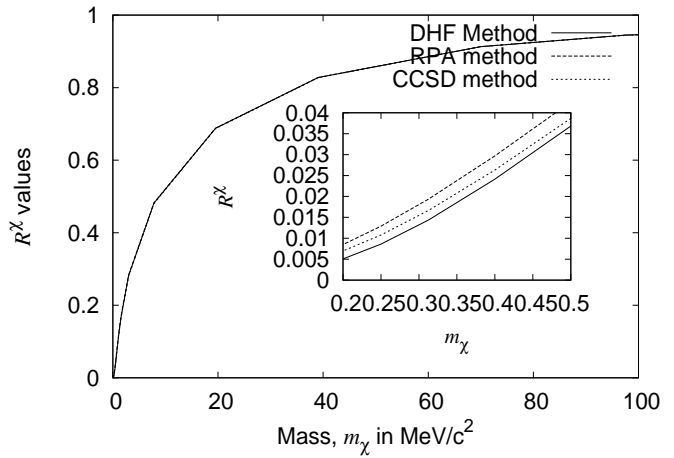} 
\caption{(Color online) Variation in the $R^{\chi}$ values with $m_{\chi}$ (in MeV/$c^2$) from the DHF, RPA and CCSD methods. We find all 
the methods are predicting almost same $R^{\chi}$ values irrespective of the fact that the trends in the $d_a^{\chi}$ results are coming out 
to be completely different way from all the three employed methods. However, slight deviations in these results can be observed for smaller
values of $m_{\chi}$ as shown in the inlet of the figure. This observation does not agree perfectly with the findings in Ref. 
\cite{derevianko} though earlier it was seen that behavior of $\rho_N^{\chi}$ was in perfect agreement between both the works.}
 \label{fig3}
 \end{figure}
 
 By combining the $\mathcal{Y}$ values from the CCSD method with the measured EDM of the $^{199}$Hg atom reported recently as 
$|d_a( ^{199}{\rm{Hg}})|< 7.4\times  10^{-30} |e|\rm{cm}$ with 95\% confidence level \cite{graner}, we get  
\begin{eqnarray}
 |C_T| < \ 7.0 \times 10^{-10} 
\end{eqnarray}
and 
\begin{eqnarray}
 |S| < 4.2 \times 10^{-13} |e| \rm{fm}^3 .
 \label{eqls}
\end{eqnarray}
Here we have used the value $\langle \sigma_N \rangle=-1/3$ of the $^{199}$Hg atomic nucleus \cite{dzuba} to deduce the limit on $C_T$. The 
above limits are comparatively larger than the estimated limits given in Ref. \cite{yashpal} owing to the fact that 
consideration of more non-linear effects in the CCSD method gives the corresponding $\mathcal{Y}$ values smaller than the values 
reported earlier.

From the relation given in Eq. (\ref{eqcp}), the S-PS e-N coupling coefficient is obtained as
\begin{eqnarray}
|C_P| < \ 1.9 \times 10^{-7}
\end{eqnarray}
and from the relation $S \simeq (1.9d_n+0.2d_p)$ fm$^2$ \cite{dmitriev}, we can get the limit on $d_n$ as
\begin{eqnarray}
 |d_n| \ < \ 2.2 \times 10^{-26} |e| \rm{cm}  
 \label{deqn}
\end{eqnarray}
against the limit $d_n < 3.0 \times 10^{-26} |e| \rm{cm}$ obtained from the direct measurement \cite{pendelbury} and on $d_p$ as
\begin{eqnarray}
 |d_p| \ < \ 2.1 \times 10^{-25} |e| \rm{cm} . 
 \label{deqp}
\end{eqnarray}
To our knowledge, these are the most reliable limits obtained on $d_n$ and $d_p$ so far.

Based on the discussion in Ref. \cite{dekens}, the tensor component contribution to $S$ from the $\pi$-n-n interactions are negligible. With
this assumption, it is given as \cite{dekens}  
\begin{eqnarray}
 S& \simeq & \big [ (0.37\pm 0.3) \bar{g}_{\pi n n}^{(0)} + (0.4 \pm 0.8) \bar{g}_{\pi n n}^{(1)} \big ] |e| \rm{fm}^3.
\end{eqnarray}
From this, we infer the limits 
\begin{eqnarray}
 |\bar{g}_{\pi n n}^{(0)}| \ < \ 1.2 \times \times 10^{-12}
\end{eqnarray}
and
\begin{eqnarray}
 |\bar{g}_{\pi n n}^{(1)}| \ < \ 1.1 \times 10^{-12} .  
\end{eqnarray}
The $\bar{\theta}$ scenario offers $\bar{g}_{\pi n n}^{(0)}=(-0.018 \pm 0.007) \bar{\theta}$ and $\bar{g}_{\pi n n}^{(1)}=(0.003 
\pm 0.002) \bar{\theta}$ \cite{dekens,vries}. This results in the best limit on $\bar{\theta}$ as \cite{dekens}    
\begin{eqnarray}
|\bar{\theta}| < \ 1.1 \times 10^{-10}.
\end{eqnarray}

Also using the limit on $\bar{g}_{\pi n n}^{(1)}$ and the relation from the the minimal supersymmetric left-right model as $\bar{g}_{\pi n n}^{(1)}
= 2 \times 10^{-12} \times (\tilde{d}_u-\tilde{d}_d)$ \cite{pospelov1,bsaisou}, we can give
\begin{eqnarray}
|\tilde{d}_u - \tilde{d}_d| \ < \ 5.5 \times 10^{-27} |e| \rm{cm} .
\end{eqnarray}
 All the above limits can be improved further when more accurate nuclear calculations on $a_i$ and experimental limit on the EDM of the
$^{199}$Hg atom are available. The SM offers an open range of value to $\bar{\theta}$ from 0 to $2\pi$, while the above restriction is 
a clear indication of existence of BSM. 

Having discussed on various limits for the infinite $m_{\chi}$ approximation, we turn onto investigating constraint on the T-PT coupling 
coefficient $C_T^{\chi}$ by considering a range of $m_{\chi}$ from the low to intermediate energy scale. In order to validate our method for 
investigating this quantity, we analyze first the behavior of $\rho_N^{\chi}(r)$ against the radial distance $r$ from the origin of the nucleus and 
compare with its trend for the given values of $m_{\chi}$ as plotted in Ref. \cite{derevianko}. These trends are shown in Fig. \ref{fig1} and we 
find that it reproduces the same trends as in Ref. \cite{derevianko}. It indicates that we have constructed $\rho_N^{\chi}(r)$ properly. Earlier,
we have also demonstrated as our RPA calculations agree with the RPA results reported in Ref. \cite{dzuba}. From this analysis, we anticipate 
to reproduce the DHF and RPA results of Ref. \cite{derevianko} and would like to demonstrate the results using the CCSD method.

\begin{figure}[t]
\includegraphics[width=9.0cm,height=6.5cm]{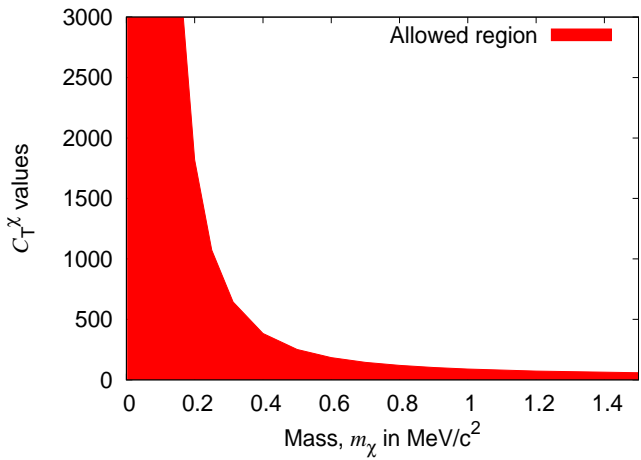}
\caption{(Color online) The area shown in color are the allowed region of the $C_T^{\chi}$ values with the $m_{\chi}$ values 
(in MeV/$c^2$). This range is substantially different than that were estimated in Ref. \cite{derevianko}.}
 \label{fig4}
 \end{figure}

 In Table \ref{tab2}, we list the $d_a^{\chi}$ values by considering a wide range of $m_{\chi}$ values from the DHF, RPA and CCSD 
methods. As can be seen, the difference between the RPA and CCSD results are quite significant for all the values of $m_{\chi}$. We have 
also given values for $m_{\chi}=\infty$. It can be noticed that these values are almost similar to that are given in Table \ref{tab1}, but the 
slight differences are owing to the consideration of the uniform nuclear charge distribution than the earlier considered Fermi nuclear 
charge distribution in the evaluation of $\rho_N^{\chi}(r)$. Again as were mentioned before due to the large cancellation between the 
all order electron core-polarization and pair-correlation effects in the CCSD method, the CCSD values are coming out to be more close 
to the DHF values than the RPA values. In fact, we highlight these differences by plotting the $d_a^{\chi}$ values for different range 
of $m_{\chi}$ values in Figs. \ref{fig2}. This plot clearly shows how RPA overestimates the electron correlation effects in the $^{199}$Hg 
atom. Thus, RPA does not seem to be suitable to employ for precise determination of constraints on the $C_T^{\chi}$ values though it accounts electron 
correlations due to core-polarization effects to all orders; rather it would be more appropriate to use results from the DHF method as an
effortless method for more valid estimate of $C_T^{\chi}$ values. Considering these results, we also determine ${\cal R^{\chi}}$ values 
from all the employed methods and plot them in Fig. \ref{fig3}. Interestingly, it can be noticed from this figure that the obtained 
${\cal R^{\chi}}$ values from all the employed methods agree quite well with each other irrespective of the fact that inclusion of the electron 
correlation effects change the $d_a^{\chi}$ values substantially from the DHF method. However, zooming the plots more closely, 
as shown in the inlet of Fig. \ref{fig3}, we find that there are small differences in the ${\cal R^{\chi}}$ values from all the considered
three methods. 

After observing that the RPA method is overestimating the electron correlation effects, we consider the CCSD results of ${\cal R^{\chi}}$
and $d_a^c/C_T$ to infer constraint on $C_T^{\chi}$. Though we find ${\cal R^{\chi}}$ values do not change much even with the inclusion of 
the electron correlation effects, but it can be seen from Table \ref{tab2} that the $C_T$ value differs significantly among the methods 
employed here. From Eq. (\ref{eqctx}), it is obvious that limit on $C_T^{\chi}$ depends on both the ${\cal R^{\chi}}$ and $C_T$ values, 
therefore it is still upholds our argument to consider a valid method to evaluate limit on $C_T^{\chi}$ more precisely. By combining these ${\cal R^{\chi}}$ and $C_T$ values from the CCSD method 
for a range of $m_{\chi}$ with the experimental limit $|d( ^{199}\rm{Hg})| < 7.4 \times 10^{-30}$ $e$cm with 95\% confidence level 
\cite{graner}, we estimate the limits on the $C_T^{\chi}$ values and plot them in Fig. \ref{fig4}. The area shown in color in this 
figure is the allowed region for the $C_T^{\chi}$ for all the considered masses of a possible dark matter candidate $\chi$. If the EDM of the 
$^{199}$Hg atom is ascertain in future then $C_T^{\chi}$ can be found out, from which we expect to determine $m_{\chi}$ value precisely. 

\section{Conclusion}

We have investigated limits on the tensor-pseudotensor and scalar-pseudoscalar coupling coefficients of the electron-nucleus interactions 
due to parity and time-reversal symmetry violations. Limits on the CP violating quantum chromodynamics parameter $\bar{\theta}$ and 
chromo-electric dipole moments of quarks were evaluated from the limits obtained on the nuclear Schiff moment. To determine these limits, 
we have combined the recently reported experimental limit on the electric dipole moment of the $^{199}$Hg atom with the sophisticated 
atomic calculations carried out based on the relativistic many-body methods. We observed from the comparison among the atomic results 
obtained by employing methods in the random phase approximation and coupled-cluster theory framework that choice of a suitable atomic 
many-body method plays crucial roles in accounting the electron correlation effects rigorously for accurate determination of the above limits at 
the cost of large computation. Further, allowing Yukawa-type parity and time-reversal symmetry violating interactions between the electrons 
and nucleus in the $^{199}$Hg atom due to a plausible dark force carrier $\chi$, constraints on the values of the tensor-pseudotensor coupling 
coefficient for a wide range of mass of $\chi$ have been imposed.

\section*{Acknowledgement}

This work was partly supported by the TDP project of Physical Research Laboratory (PRL), Ahmedabad and the computations were carried out 
using the Vikram-100 HPC cluster of PRL.

\end{document}